\documentclass[twocolumn,showpacs,amsfonts,aps,prl,floatfix,%
superscriptaddress]{revtex4} 

\usepackage{amsmath}
\usepackage{bm}
\usepackage{graphicx}

\newcommand{\Tc}{T_\mathrm{c}}

\newcommand{\be}[1]{\begin{equation}\label{#1}}
\newcommand{\ee}{\end{equation}}

\newcommand{\eq}{{\,=\,}}

\begin{document}


\title{Suppression of elliptic flow in a minimally viscous quark-gluon 
plasma}
\date{\today}

\author{Huichao Song}
\affiliation{Department of Physics, The Ohio State University, 
  Columbus, OH 43210, USA}
\author{Ulrich Heinz}
\email[Correspond to\ ]{heinz@mps.ohio-state.edu}
\affiliation{Department of Physics, The Ohio State University, 
  Columbus, OH 43210, USA}
\affiliation{CERN, Physics Department, Theory Division, CH-1211 Geneva 23, 
Switzerland}

\begin{abstract}
We compute the time evolution of elliptic flow in non-central relativistic 
heavy-ion collisions, using a (2+1)-dimensional code with longitudinal
boost-invariance to simulate viscous fluid dynamics in the causal 
Israel-Stewart formulation. We show that even ``minimal'' shear 
viscosity $\eta/s=\hbar/(4\pi)$ leads to a large reduction of elliptic flow
compared to ideal fluid dynamics, raising questions about the interpretation 
of recent experimental data from the Relativistic Heavy Ion Collider.     
\end{abstract}


\maketitle

The success of the hydrodynamic model in describing the bulk of 
hadron production in Au+Au collisions at the Relativistic Heavy
Ion Collider (RHIC) \cite{reviews} has led to a paradigmatic shift
in our view of the quark-gluon plasma (QGP): Instead of behaving
like a gas of weakly interacting quarks and gluons \cite{QGP}, as 
naively expected on the basis of asymptotic freedom in QCD, its
collective properties rather reflect those of a ``perfect liquid''
with (almost) vanishing viscosity. However, due to quantum mechanical
uncertainty no fluid can have exactly zero viscosity 
\cite{Danielewicz:1984ww}, and recent work \cite{Policastro:2001yc}
on strongly coupled gauge field theories, based on techniques exploiting 
the AdS/CFT correspondence, suggests an absolute lower limit for the
ratio of shear viscosity $\eta$ to entropy density $s$: 
$\eta/s\geq\hbar/4\pi$. This raises the interesting question how
close to this limit the actual value of the shear viscosity of the QGP
created at RHIC is. 

Answering this question requires hydrodynamic 
simu\-lations for relativistic {\em viscous} fluids in which the ratio
$\eta/s$ enters as a parameter. To study the anisotropic (``elliptic'') 
collective flow in non-central heavy-ion collisions, from which limits 
on $\eta/s$ can be extracted \cite{Teaney:2003kp}, requires a code that
evolves the hydrodynamic fields at least in the two dimensions transverse
to the heavy-ion beam. In this Letter we present our first results from such 
simulations \cite{perimeter}; a longer paper with a discussion of all 
technical details of our approach is in preparation \cite{long}.

Relativistic hydrodynamics of viscous fluids is technically demanding.
The straightforward relativistic generalization of the non-relativistic
Navier-Stokes equation yields unstable equations that can lead
to acausal signal propagation. A causally consistent theoretical 
framework was developed 30 years ago by Israel and Stewart 
\cite{Israel:1976tn}. It involves the simultaneous solution of 
hydrodynamic equations for a generalized energy-momentum tensor containing
viscous pressure contributions, $\pi^{\mu\nu}(x)$, together with kinetic 
evolution equations, characterized by a (short) microscopic collision 
time scale $\tau_\pi$, for the dynamical approach of $\pi^{\mu\nu}$ 
towards its Navier-Stokes limit. Compared to ideal fluid dynamics, this 
leads effectively to more than a doubling of the number of coupled partial 
differential equations to be solved \cite{Heinz:2005bw}.

The last couple of years have seen extensive activity in implementing the 
Israel-Stewart equations (and slight variations thereof) 
\cite{Israel:1976tn,Heinz:2005bw,Muronga:2001zk,Teaney:2004qa,Baier:2006um}
numerically, for systems with boost-invariant longitudinal expansion
and transverse expansion in zero \cite{Muronga:2001zk,Baier:2006um},
one \cite{Teaney:2004qa,Muronga:2004sf,Chaudhuri:2005ea,Baier:2006gy}
and two dimensions \cite{Chaudhuri:2007zm,Romatschke:2007mq}
(see also Ref.~\cite{Chaudhuri:2006jd} for a numerical study of the 
relativistic Navier-Stokes equation in 2+1 dimensions). It is probably
fair to say that the process of verification and validation of these
numerical codes is still ongoing: While different initial conditions and 
evolution parameters used by the different groups of authors render a 
direct comparison of their results difficult, it seems unlikely that 
accounting for these differences will bring the various published 
numerical results in line with each other. 

We here present results obtained with an independently developed 
(2+1)-dimensional causal viscous hydrodynamic code, VISH2+1 
\cite{perimeter,fn1}, which has been extensively tested (for details 
see \cite{long}): (i) in the limit of vanishing viscosity, it accurately 
reproduces results obtained with the (2+1)-d ideal fluid code AZHYDRO 
\cite{AZHYDRO}; (ii) for homogeneous density distributions (i.e. in the 
absence of density gradients) and vanishing relaxation time it accurately 
reproduces the known analytic solution of the relativistic Navier-Stokes 
equation for boost-invariant longitudinal expansion 
\cite{Danielewicz:1984ww}; (iii) for very short kinetic relaxation times 
our Israel-Stewart code accurately reproduces results from a (2+1)-d 
relativistic Navier-Stokes code, under restrictive conditions where the 
latter produces numerically stable solutions; and (iv) for simple 
analytically parametrized anisotropic velocity profiles the numerical 
code correctly computes the velocity shear tensor that drives the 
viscous hydrodynamic effects. 

VISH2+1 solves the equations for local energy-momentum conservation,
$d_m T^{mn}\eq0$, with
%
\be{Tmunu} 
 T^{mn} = e u^m u^n - p\Delta^{mn} + \pi^{mn},\ 
 \Delta^{mn} = g^{mn}{-}u^m u^n,
\ee
%
together with kinetic equations for the viscous shear pressure 
$\pi^{mn}$ \cite{fn2},   
\be{pikin}
  D\pi^{mn} = 
  \frac{1}{\tau_{\pi}}(2\eta\sigma^{mn}{-}\pi^{mn})
  -(u^m\pi^{nk}{+}u^n\pi^{mk}) Du_k.
\ee
Here $D{\eq}u^m d_m$ is the time derivative in the local
comoving frame and $\sigma^{mn}\eq\nabla^{\left\langle m\right.}
u^{\left.n\right\rangle}\eq\frac{1}{2}(\nabla^m u^n{+}\nabla^n
u^m)-\frac{1}{3}\Delta^{mn}d_k u^k$ (with $\nabla^m\eq\Delta^{ml} d_{l}$) 
is the symmetric and traceless velocity shear tensor. 
We use a fixed Eulerian space-time grid in curvilinear coordinates 
$x^m\eq(\tau,x,y,\eta)$, with longitudinal proper time 
$\tau\eq\sqrt{t^2{-}z^2}$ and space-time rapidity 
$\eta\eq\frac{1}{2}\ln\frac{t{+}z}{t{-}z}$, where $z$ is the beam direction 
and $(x,y)$ are the two transverse directions. $d_m$ indicates the covariant 
derivative in direction $x^m$ in this coordinate system. As in 
Refs.~\cite{Chaudhuri:2005ea,Baier:2006gy,Chaudhuri:2007zm,Romatschke:2007mq,%
Chaudhuri:2006jd} we neglect bulk viscosity and heat conduction as presumably 
subdominant effects in a QGP with approximately vanishing net baryon density.

We implement longitudinal boost-invariance via the ansatz 
$u^m\eq(u^\tau,u^x,u^y,u^\eta)\eq\gamma_\perp (1,v_x,v_y,0)$, with 
$\gamma_\perp=(1{-}v_x^2{-}v_y^2)^{-1/2}$, using $\eta$-independent 
initial conditions. The equations to be solved are 3 hydrodynamic 
equations for $T^{\tau\tau}$,  $T^{\tau x}$ and $T^{\tau y}$, together
with 4 kinetic equations for $\pi^{\eta\eta}$, $\pi^{\tau\tau}$, 
$\pi^{\tau x}$ and $\pi^{\tau y}$ \cite{long,fn3}. To check the numerics 
we also evolved additional, redundant components of $\pi^{mn}$ and 
confirmed that the identities $u_m\pi^{mn}\eq0\eq\pi^m_m$ are preserved
over time.

Due to the limited size of the transverse $(x,y)$ grid in our current code, 
we here present results only for Cu+Cu collisions; simulations of 
the larger Au+Au system will soon be forthcoming. We
use standard Glauber model initial conditions \cite{reviews}, assuming
wounded-nucleon scaling of the initial transverse energy density 
$e(x,y,\tau_0;b)$ \cite{Kolb:2000sd}, with a Woods-Saxon radius 
$R_\mathrm{Cu}=4.2$\,fm, surface thickness $\xi\eq0.596$\,fm, and
equilibrium density $\rho_0\eq0.17$\,fm$^{-3}$. We scale this profile
to a peak initial energy density in central $(b\eq0)$ collisions of 
$e_0\eq30$\,GeV/fm$^3$. This is higher than expected for Cu+Cu collisions
at RHIC but ensures that the system spends enough time in the QGP phase
to explore the effects of shear viscosity on the evolution of anisotropic
flow in this phase. We start the hydrodynamic evolution at 
$\tau_0\eq0.6$\,fm/$c$ with vanishing transverse flow, both 
for the viscous evolution and ideal fluid dynamical comparison runs.
For the equation of state (EoS) we use a slight variation of EOS~Q from 
Ref.~\cite{Kolb:2000sd} (which implements a phase transition at 
$\Tc\eq164$\,MeV between a free quark-gluon gas above $\Tc$ and a 
chemically equilibrated hadron resonance gas below $\Tc$) where the sharp 
corners at either end of the Maxwell contruction at $\Tc$ are rounded off
for numerical stability \cite{fn3a}. Neither the initial conditions nor 
the EoS have been fine-tuned for a realistic comparison with experimental 
data; we here emphasize the comparison between ideal fluid and viscous 
hydrodynamic evolution, in order to identify qualitative differences 
between the two and to quantitatively understand their origin.

The viscous hydrodynamic equations contain two parameters, the shear
viscosity $\eta$ and the kinetic relaxation time $\tau_\pi$. All simulations
presented here are done with ``minimal'' viscosity \cite{Policastro:2001yc}
$\eta/s=1/4\pi$, while $\tau_\pi$ is varied between $\frac{1}{2}$ 
(default) and $\frac{1}{4}$ of the classical Boltzmann gas estimate 
$\tau_\pi^\mathrm{Boltz}\eq\frac{6\eta}{4p}\eq\frac{6}{T}\frac{\eta}{s}$
\cite{Israel:1976tn,fn4}. 

%
\begin{figure}[t]
\includegraphics[width =0.98\linewidth,clip=]{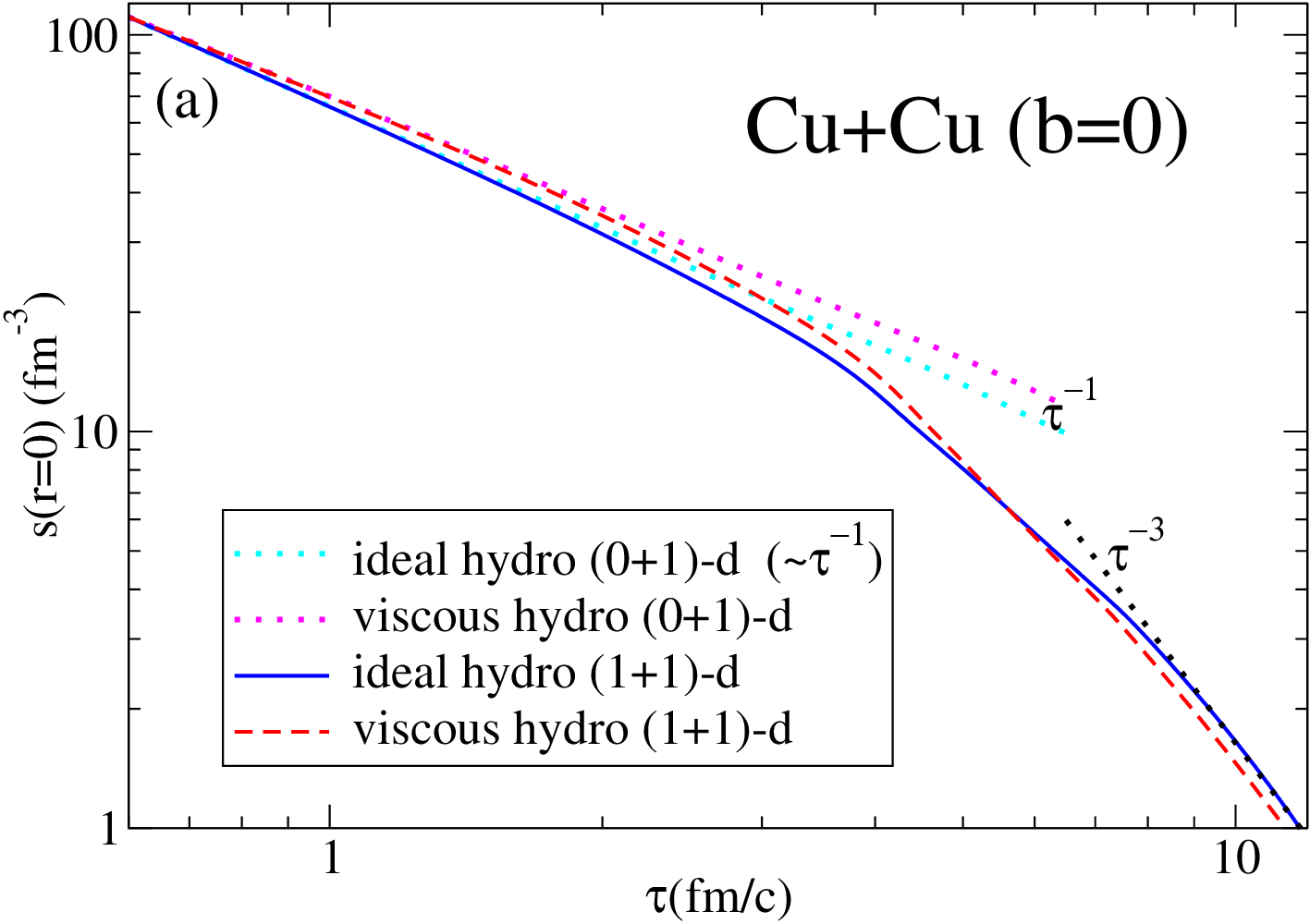}\\
\includegraphics[width =0.98\linewidth,clip=]{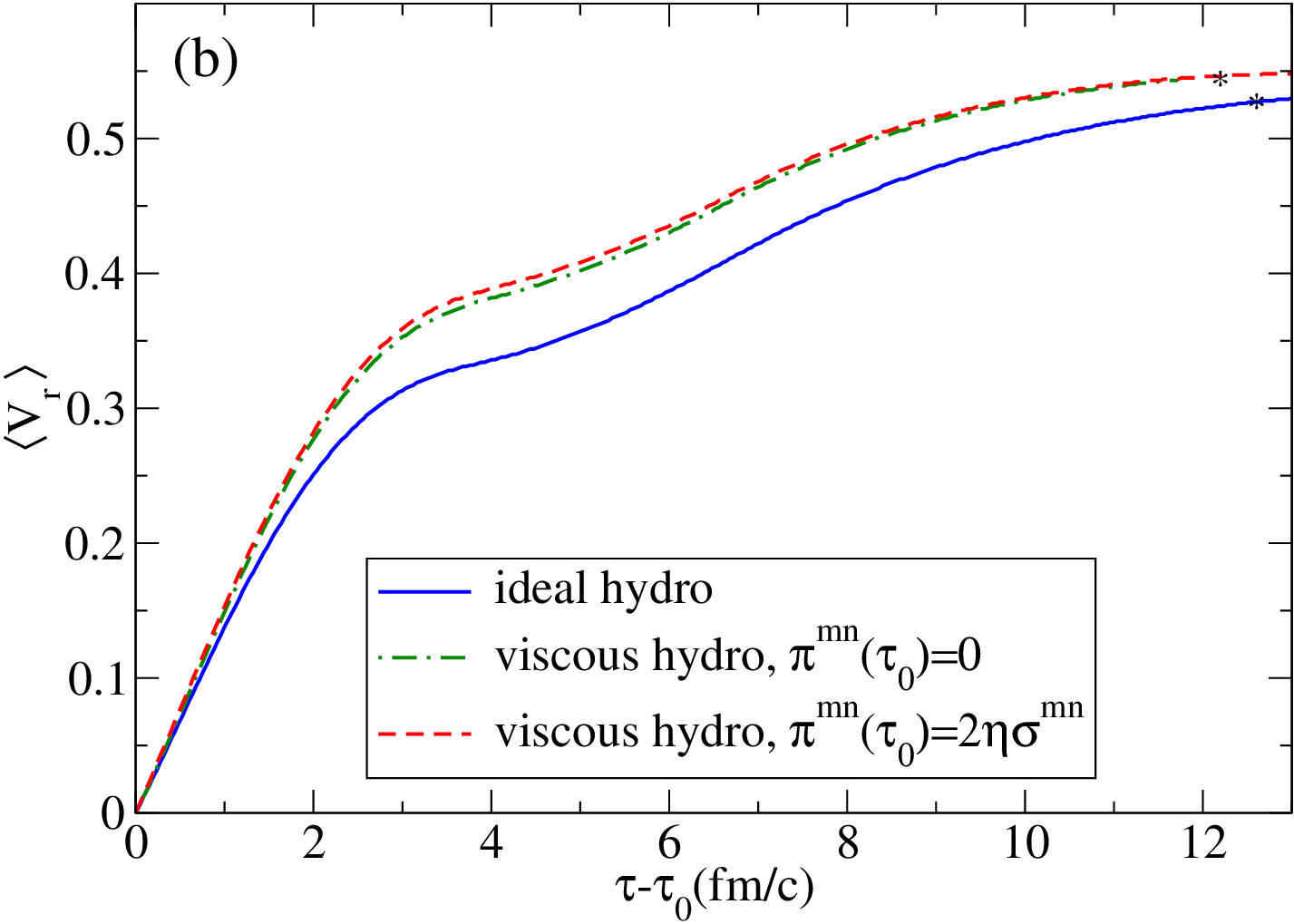}
\caption{(Color online)
Time evolution of the central entropy density (a) and average
radial flow velocity (b) in central Cu+Cu collisions (see text for details).
Here and later stars indicate the time when all matter is frozen out.}
\label{F1}
\end{figure}
%

In Figure~\ref{F1} we show the evolution of the central entropy density 
$s(r{=}0)$ and the average radial flow $\langle v_r\rangle$ (with the 
Lorentz contracted energy density $\gamma_\perp e$ as weight function)
for central $(b{=}0)$ Cu+Cu collisions. The curves labeled ``(0+1)-d'' 
correspond to 1-dimensional boost-invariant longitudinal expansion 
without transverse flow (i.e. to transversally homogeneous initial 
conditions). One sees that in this case (which, for a simple EoS
and in the Navier-Stokes limit, can be solved analytically 
\cite{Danielewicz:1984ww}) shear viscosity reduces the cooling 
rate, due to a well-known {\em reduction of the work done by longitudinal 
pressure} \cite{Danielewicz:1984ww} -- additional entropy is produced, 
and the entropy density decreases more slowly than the $1/\tau$-law for 
ideal fluids. For transversally inhomogeneous initial conditions 
(``(1+1)-d hydro''), the developing radial flow increases the cooling 
rate compared to the case without transverse expansion, for both ideal 
and viscous fluid dynamics. However, in this case the shear viscous 
effects {\em increase the transverse pressure} relative to the ideal 
fluid case, leading to a faster build-up of radial flow (bottom panel 
in Fig.~\ref{F1}), which in turn accelerates the cooling of the fireball 
center in such a way that by the time the expansion becomes effectively 
three-dimensional (indicated by the $\tau^{-3}$ line in Fig.~\ref{F1}(a))
the viscous fluid cools {\em more rapidly} than the ideal one 
\cite{Teaney:2004qa}. (This was also seen (although not emphasized) in 
Refs.~\cite{Teaney:2004qa,Chaudhuri:2005ea,Baier:2006gy,Chaudhuri:2007zm}.) 
The larger radial flow of the viscous fluid (Fig~\ref{F1}(b)) leads to 
flatter final transverse momentum spectra \cite{Heinz:2002rs,Chaudhuri:2005ea,%
Baier:2006gy,Chaudhuri:2007zm}, but (as pointed out in 
\cite{Heinz:2002rs,Baier:2006gy}) this effect can be largely compensated by 
starting the viscous hydrodynamic evolution later and with lower initial 
energy density. 

%
\begin{figure}[htb]
\includegraphics[width =0.98\linewidth,clip=]{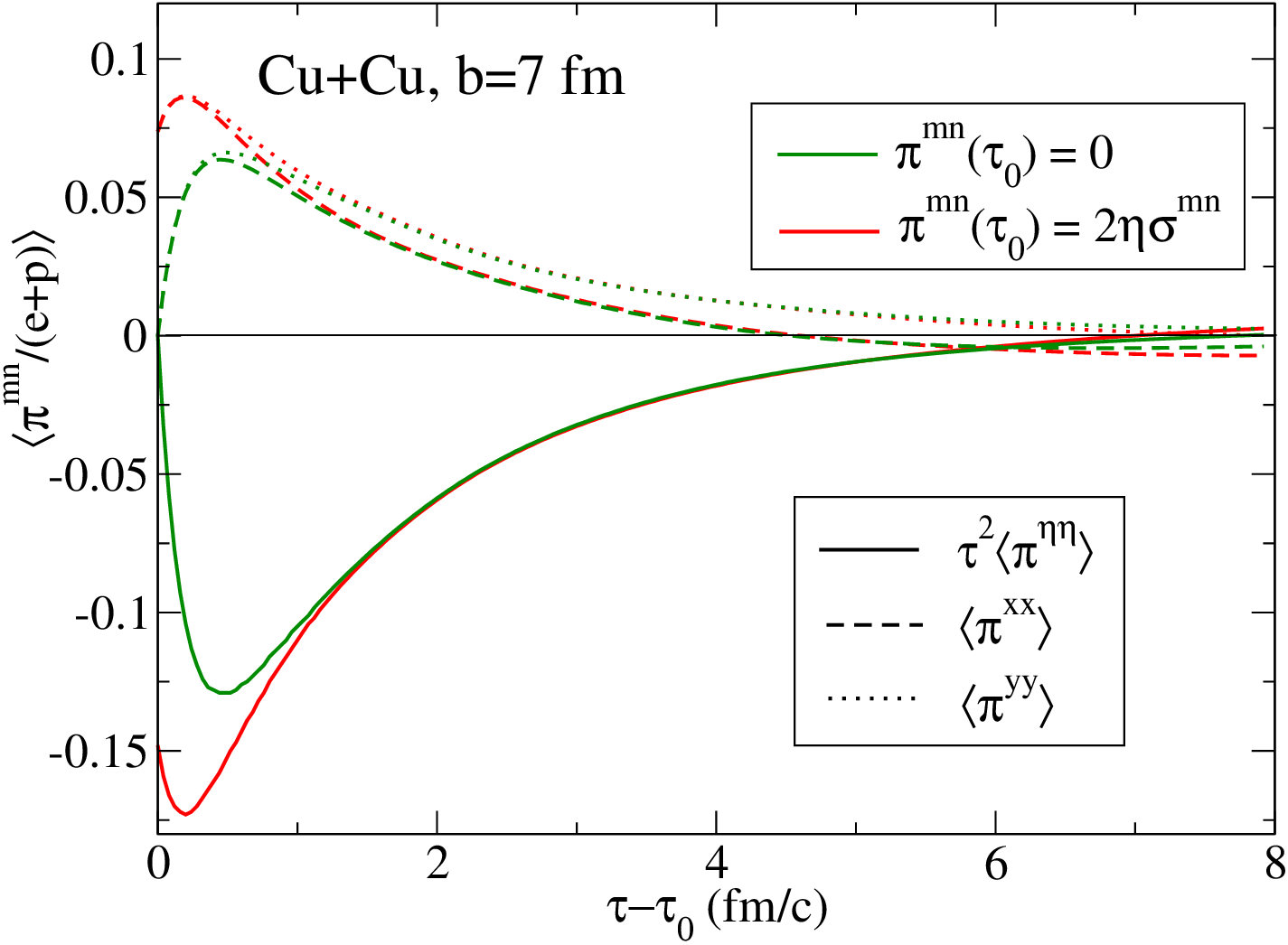}
\caption{(Color online) Time evolution of the dominant components of the 
shear viscous pressure tensor, normalized by $e{+}p$ and averaged over 
the transverse plane, for two different initial conditions (red and green, 
respectively). Note that the normalization factor $e{+}p{\,\sim\,}T^4$ 
decreases rapidly with time.
}
\label{F2}
\end{figure}
%
The dotted and dash-dotted lines in Fig.\,\ref{F1}(b) show that (at least 
for the short relaxation time $\tau_\pi\eq\frac{1}{2}\tau_\pi^\mathrm{Boltz}
=0.24\left(\frac{200\,\mathrm{MeV}}{T}\right)$\,fm/$c$ considered here) 
the initial conditions for the shear viscous pressure tensor don't matter 
much: whether $\pi^{mn}$ is initially taken to vanish or to assume its 
Navier-Stokes limit $2\eta\sigma^{mn}$, it quickly relaxes to the same
function at times $\tau{-}\tau_0 \gtrsim4\tau_\pi\approx 1$\,fm/$c$.
This is seen more explicitly in Fig~\ref{F2} where we show, for
non-central Cu+Cu collisions at $b\eq7$\,fm and the same two sets of
initial conditions for the viscous shear pressure tensor, the time 
evolution of the dominant components of $\pi^{mn}$, normalized to the 
equilibrium enthalpy $e{+}p$ (which sets the scale for the energy-momentum
tensor in the ideal fluid limit) and averaged over the transverse plane. 
Other components of $\langle\pi^{mn}\rangle$ are at least an order of 
magnitude smaller than the ones shown. The signs of 
$\langle\pi^{\eta\eta}\rangle<0$ and $\langle\pi^{xx}\rangle,
\langle\pi^{yy}\rangle>0$ reflect the reduced longitudinal and increased 
transverse pressure caused by shear viscosity. The negative difference 
$\langle(\pi^{xx}{-}\pi^{yy})/(e{+}p)\rangle<0$ seen in Fig.~\ref{F2} 
causes a significant viscous reduction of the total momentum anisotropy 
$\epsilon_p$ which we discuss next.

%
\begin{figure}[htb]
\includegraphics[width =0.98\linewidth,clip=]{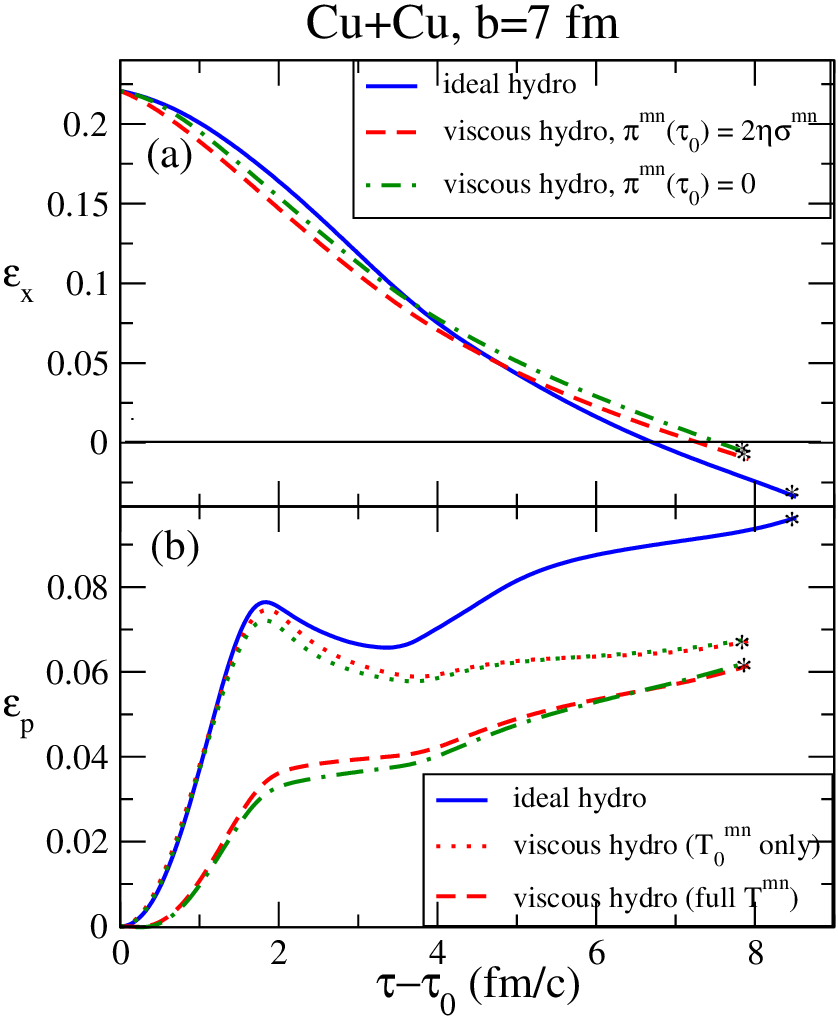}
\caption{(Color online) Time evolution of the spatial eccentricity 
$\epsilon_x$ (a) and momentum anisotropy $\epsilon_p$ (b) for non-central 
Cu+Cu collisions at $b\eq7$\,fm. See text for details. 
}
\label{F3}
\end{figure}
%

The anisotropic flow is driven by the spatial source eccentricity
$\epsilon_x\eq\frac{\langle y^2{-}x^2\rangle}{\langle y^2{+}x^2\rangle}$
(where $\langle\dots\rangle$ denotes the energy density weighted
average over the transverse plane \cite{reviews}) and the anisotropic
pressure gradients it generates. Fig.~\ref{F3}a shows that it
decreases as a function of time, and does so more rapidly initially, but
more slowly later for the viscous fluid than in ideal hydrodynamics.
The initial drop rate for $\epsilon_x$ depends on the initial value
for the viscous pressure tensor, but after about 1\,fm/$c$ different
initializations for $\pi^{mn}$ lead to parallel evolution histories
for $\epsilon_x$. The largest initial decrease for $\epsilon_x$ is 
observed for the largest initial viscous pressure tensor; initial 
free-streaming of the matter would correspond to an extreme case
of viscous fluid dynamics, leading to even larger initial $\pi^{mn}$
and even faster initial decrease of the spatial eccentricity than
shown in Fig.~\ref{F3}a \cite{HHunpub}.

In Fig.~\ref{F3}b we plot, for ideal and viscous hydrodynamic evolution, 
the total momentum anisotropy averaged over the transverse plane,
$\epsilon_p\eq\frac{\langle T^{xx}{-}T^{yy}\rangle}
                   {\langle T^{xx}{+}T^{yy}\rangle}
= \frac{\langle T_0^{xx}{+}\pi^{xx}-T_0^{yy}{-}\pi^{yy}\rangle}
       {\langle T_0^{xx}{+}\pi^{xx}+T_0^{yy}{+}\pi^{yy}\rangle}$,
as a function of time. The closely spaced green and red lines
distinguish different initializations for $\pi^{mn}$, as specified
in Fig.~\ref{F3}a, showing weak sensitivity to these initial values.
More interesting is the separation of the contributions to $\epsilon_p$
arising from the ideal fluid part $T_0^{mn}$ (which only tracks the
differences in the evolution of flow velocity and thermal pressure between
ideal and viscous hydrodynamics) and from the viscous pressure components
$\pi^{mn}$ (dashed and dotted lines in Fig.~\ref{F3}b). Initially, the 
viscous flow and thermal pressure evolution is not very different from the
ideal fluid case, and it takes a while until the viscous pressure effects
manifest themselves in a significant reduction of the flow anisotropy,
thereby modifying the ideal fluid part $T_0^{mn}$ of the energy momentum 
tensor. However, the viscous pressure components $\pi^{mn}$
contribute themselves a negative, initially large part to the total momentum 
anisotropy, resulting (for $\frac{\eta}{s}\eq\frac{1}{4\pi}$) in an overall 
reduction of the latter by almost 50\% relative to the ideal fluid case 
{\em over the entire time history}. At late times, the viscous pressure
components become small, but by then their negative effect on the buildup
of elliptic flow has fully manifested itself in the collective flow
profile and is thus carried by the ideal fluid part $T_0^{mn}$ of
the energy momentum tensor. 

%
\begin{figure}[htb]
\includegraphics[width =0.98\linewidth,clip=]{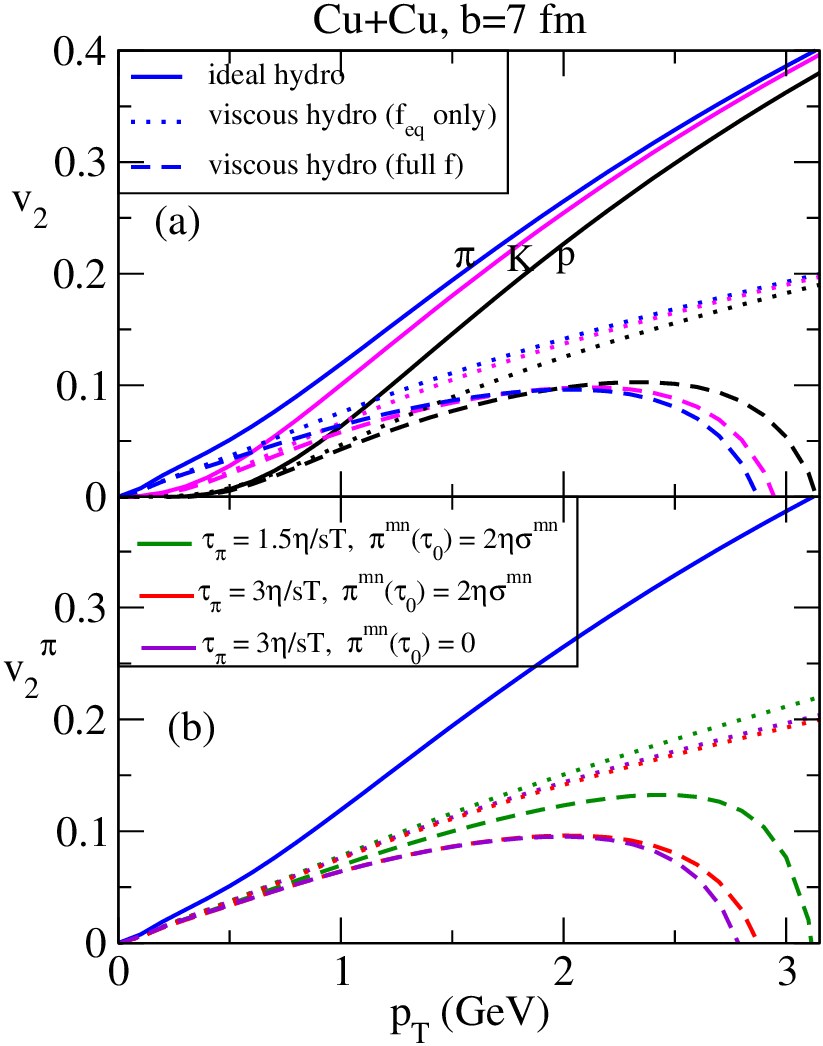}\\
\caption{(Color online) 
(a) The elliptic flow $v_2(p_T)$ for pions, kaons and protons
from ideal fluid dynamics (solid lines) and viscous hydrodynamics
(dotted and dashed lines). Dotted lines account only for viscous 
effects on the flow pattern that enters the equilibrium part of the 
distribution function; dashed lines additionally include viscous 
(non-equilibrium) corrections to the latter. (b) Effects of
different choices for the kinetic relaxation time $\tau_\pi$ and for
the initial viscous pressure $\pi^{mn}$ on pion elliptic flow (with 
the same separation of equilibrium and non-equilibrium contributions 
as in part (a)).  
}
\label{F4}
\end{figure}
%

The elliptic flow $v_2^{(i)}\eq\langle\cos(2\phi_p)\rangle_i$, calculated 
as the $\cos(2\phi_p)$-moment of the observed hadron momentum distri\-bution
$\frac{dN_i}{dy\,p_Tdp_T\,d\phi_p}$, depends on particle species $i$ and 
measures the distribution of the total momentum anisotropy $\epsilon_p$ over 
the various hadron species and over transverse momentum $p_T$. This 
distribution depends on the chemical composition of the fireball at
kinetic freeze-out, here assumed (unrealistically) to be a chemically 
equilibrated hadron resonance gas. The hadron momentum distributions
are calculated as a Cooper-Frye integral \cite{reviews} over a decoupling
surface $\Sigma$ of constant temperature $T_\mathrm{dec}\eq130$\,MeV:
\begin{eqnarray}
\label{fo}
  &&\frac{dN_i}{dy p_Tdp_T d\phi_p} =
  \int_\Sigma\!\! \frac{p\cdot d^3\sigma(x)}{(2\pi)^3}
  \left[f_\mathrm{eq}^{(i)}(x,p)+\delta f^{(i)}(x,p)\right]
\nonumber\\ 
  &&= \int_\Sigma \frac{p\cdot d^3\sigma(x)}{(2\pi)^3}
      \, f_\mathrm{eq}^{(i)}(x,p)
      \\ \nonumber
  && \qquad\qquad \times
      \left[1 + \bigl(1{-}f_\mathrm{eq}^{(i)}(x,p)\bigr)
                \frac{1}{2} \frac{p^mp^n}{T^2(x)} 
                         \frac{\pi_{mn}(x)}{(e{+}p)(x)}\right].
\end{eqnarray}
Here $f_\mathrm{eq}^{(i)}\!\left(\frac{p\cdot u(x)}{T(x)}\right)$ is the 
local thermal equilibrium distribution function with temperature $T(x)$, 
boosted to the laboratory frame by the local hydrodynamic flow velocity 
$u^m(x)$ -- both 
taken from the hydrodynamic output 
on the freeze-out surface $\Sigma$ (whose normal vector is $d^3\sigma(x)$).
$\delta f^{(i)}$ describes the deviation from local thermal equilibrium
due to viscous effects and is given by the last term in the bottom line
of Eq.~(\ref{fo}) \cite{Teaney:2003kp,Baier:2006um}. It is proportional
to the viscous pressure tensor $\pi^{mn}$, and even though (in 
contrast to early times) $\pi^{mn}(x)$ is small on the freeze-out surface, 
its effect on the distribution function grows quadratically with momentum,
leading to a breakdown of the (viscous) hydrodynamic calculation of 
hadron spectra at sufficiently large $p_T$ (in our case 
$|\delta N^\pi(p)/N^\pi_\mathrm{eq}(p)|{\,>\,}50\%$ for 
$p_T{\,>\,}2.5$\,GeV/$c$ where $N^\pi(p)$ denotes the pion momentum 
spectrum).

In Fig.~\ref{F4} we compare, for a few common hadron species, the elliptic 
flow $v_2(p_T)$ from ideal and viscous fluid dynamics.
One sees that even ``minimal'' shear viscosity $\frac{\eta}{s}\eq\frac{1}{4\pi}$
causes a dramatic suppression of elliptic flow. The effects seen in 
Fig.~\ref{F4} seem to be even larger than those reported in 
\cite{Romatschke:2007mq} -- this discrepancy clearly calls for clarification
\cite{fn5}. Even without accounting for the slight deviations $\delta f$ of 
the distribution function from its thermal equilibrium form, caused by small 
but non-vanishing shear pressure tensor components on the freeze-out surface, 
we see that elliptic flow is reduced by 
almost 50\% (dotted lines in Fig.~\ref{F4}), due to the already mentioned 
reduction of azimuthal anisotropies of the hydrodynamic flow field. This effect
was not even considered by Teaney in \cite{Teaney:2003kp} when he attempted
to constrain the shear viscosity of the QGP using RHIC $v_2$ data. Teaney's
argument was entirely based on the viscous corrections 
$\delta f\sim p^mp^n\pi_{mn}\sim p_T^2$ arising from non-vanishing viscous 
pressure on the freeze-out surface (the difference between the dotted and 
dashed lines in Fig.~\ref{F4}) which, at low $p_T$, is a much smaller 
effect. His phenomenological limit \cite{Teaney:2003kp} on the shear 
viscosity of the fireball matter created at RHIC is therefore not 
restrictive enough: even a superficial comparison of the shapes of
the $v_2(p_T)$ curves in Fig.~\ref{F4} with experimental data 
\cite{reviews} suggests that {\em RHIC data may be inconsistent with 
the conjectured lower limit $\frac{\eta}{s}\eq\frac{1}{4\pi}$ for the QGP 
shear viscosity}. 

This conclusion, even though tentative since it is not
yet based on a quantitative data comparison with calculations that use 
a more realistic EoS and better initial conditions, appears to be robust 
since neither a 100\% variation of the initial value for $\pi^{mn}$ nor a 
50\% reduction of the kinetic relaxation time for the viscous pressure
tensor (see Fig.~\ref{F4}b) are able to significantly attenuate the 
strong viscous reduction of elliptic flow that we see in our calculations.
While it supports the new paradigm of the ``perfect fluidity'' of the
QGP created at RHIC, a possible violation of the conjectured ``minimum 
viscosity bound'' \cite{Policastro:2001yc} is a serious matter that
raises the question whether other explanations might be possible. 
Among the possibilities that one might contemplate are that the initial
spatial source eccentricity in RHIC collisions has been significantly 
underestimated \cite{Hirano:2005xf} or that a realistic equation of 
state is effectively stiffer than the one used here. T.~Cohen and 
collaborators \cite{Cohen:2007qr} have advanced counter examples of 
theories that appear to contradict the existence of a universal 
``minimum viscosity bound'', although these examples to not include QCD.
Our findings suggest that QCD might belong to the list of exceptions.
Lublinsky and Shuryak \cite{Lublinsky:2007mm} have shown that higher
order corrections to the Israel-Stewart theory of viscous relativistic
hydrodynamics tend to decrease viscous entropy production, so there may
be a possibility that they also reduce viscous effects on the elliptic 
flow \cite{fn5}. If this turns out to be case, our tentative conclusion 
as formulated above is premature. Obviously all these issues must be 
carefully investigated before a complete understanding of the RHIC data 
can be achieved.    
 
This work was supported by the U.S. Department of Energy under contract 
DE-FG02-01ER41190.



\begin{thebibliography}{99}
\bibitem{fn3a}
  Specifically, we smooth the discontinuities of the squared speed of 
  sound $c_s^2$ (which vanishes in the mixed phase) at both ends of
  the mixed phase with a Fermi function, with width parameters 
  $\Delta e\eq0.1$\,GeV/fm$^3$ at the upper (QGP) end of the mixed phase
  and $\Delta e\eq0.02$\,GeV/fm$^3$ at its lower (HG) end.
\end{thebibliography}

\end{document}